# Neutron scattering study of the kagome metal $Sc_3Mn_3Al_7Si_5$

X. Y. Li[1,2*], D. Reig-i-Plessis[1], P. -F. Liu[3], S. Wu[4,5], B. -T. Wang[3,6], A. M. Hallas[1], M. B. Stone[7], C. Broholm[8], M. C. Aronson[1]

[1]*Department of Physics & Astronomy and Stewart Blusson Quantum Matter Institute, University of British Columbia, Vancouver V6T 1Z4, Canada*

[2]*Department of Physics and Astronomy, Texas A&M University, College Station, Texas 77843, USA*

[3]*Institute of High Energy Physics, Chinese Academy of Sciences, Beijing 100049, China*

[4]*Department of Physics, University of California Berkeley, Berkeley, California 94720, USA*

[5]*Material Sciences Division, Lawrence Berkeley National Lab, Berkeley, California, 94720, USA*

[6]*Collaborative Innovation Center of Extreme Optics, Shanxi University, Taiyuan 030006, China*

[7]*Neutron Scattering Division, Oak Ridge National Laboratory, Oak Ridge, Tennessee 37831, USA*

[8]*Institute for Quantum Matter and Department of Physics and Astronomy, The Johns Hopkins University, Baltimore, Maryland 21218, USA*

**ABSTRACT:**

$Sc_3Mn_3Al_7Si_5$ is a rare example of a correlated metal in which the Mn moments form a kagome lattice. The absence of magnetic ordering to the lowest temperatures suggests that geometrical frustration of magnetic interactions may lead to strong magnetic fluctuations. We have performed inelastic neutron scattering measurements on $Sc_3Mn_3Al_7Si_5$, finding that phonon scattering dominates for energies from ~20 - 50 meV. These results are in good agreement with *ab initio* calculations of the phonon dispersions and densities of states, and as well reproduce the measured specific heat. A weak magnetic signal was detected at energies less than ~10 meV, present only at the lowest temperatures. The magnetic signal is broad and quasielastic, as expected for metallic paramagnets.

*To whom correspondence should be addressed. xiyang.li@ubc.ca

I. INTRODUCTION

A quantum spin liquid (QSL) is an exotic state of matter with fractionalized excitations and long-range entanglement, but no symmetry breaking [1]. Although no one material is universally accepted to have a QSL ground state, there are a number of QSL candidates. Most are insulating [2–5], where spatially localized magnetic moments are arranged on lattices with triangular motifs, leading to geometrical frustration and strong quantum fluctuations that prevent magnetic order [5–9]. Recently, metallic systems have been discovered where magnetic moments are at least partially itinerant, and where new quantum ground states might become possible [10–16]. We focus here on $Sc_3Mn_3Al_7Si_5$, a rare example of a metallic system where Mn atoms form well-separated and undistorted kagome planes [10].

Measurements of the magnetic susceptibility and specific heat to temperatures as low as $T$=1.8 K find no evidence of magnetic order in $Sc_3Mn_3Al_7Si_5$ [10]. Above ~50 K, the magnetic susceptibility is described by a Curie-Weiss expression, with a moment of 0.5 $\mu_B$/Mn that is much suppressed relative to Hund's rule values. This indicates that the magnetism in $Sc_3Mn_3Al_7Si_5$ has a pronounced itinerant character. The Weiss temperature $\theta_W$=-38 K [10] sets the scale for possible magnetic order, and the absence of order at temperatures as low as 1.8 K indicates that strong quantum fluctuations likely suppress that order to zero temperature. The geometrical frustration of the kagome lattice is a natural source of quantum fluctuations, and evidence for their presence in $Sc_3Mn_3Al_7Si_5$ is found in the low temperature magnetic susceptibility, which diverges rapidly below 20 K, and the specific heat, which displays a weak power law in $C_p/T$ below ~10 K [10].

Neutron scattering is the probe of choice for studying magnetic fluctuations. We report here the results of inelastic neutron scattering (INS) measurements that seek evidence for the quantum critical magnetic excitations that are implied by the bulk measurements, and in particular the continuum of fractionalized magnetic excitations that are expected in kagome systems hosting a QSL ground state [1,5,7,8]. We find magnetic excitations with weak scattering intensity in $Sc_3Mn_3Al_7Si_5$, only detected at the lowest energies and wave vectors, being masked by strong phonon scattering over much of the range of energies and wave vectors that are accessed in this experiment. Our measurements show that the measured phonon density of states (DOS) is in good agreement with computational results, and as well accounts quantitatively for the phonon contribution to the measured specific heat.

## II. METHODOLOGY

### A. Crystal growth and structure characterization

Bar-shaped crystals of $Sc_3Mn_3Al_7Si_5$ were grown from a reactive aluminum flux [10]. Single crystal X-ray diffraction (XRD) measurements confirmed that $Sc_3Mn_3Al_7Si_5$ is described by the hexagonal *P*6$_3$/*mmc* (No.194) space group, where Mn atoms form kagome planes (Fig. 1) in which the Mn-Mn spacing is 4.1759 Å. These kagome planes are separated by Sc/Al/Si slabs resulting in an interplanar Mn-Mn spacing of 4.5422 Å [10].

Single crystal XRD measurements were performed for temperatures between 100 K and 240 K with a step increment of 20 K. The reported crystal structure was verified at all temperatures, and a volumetric thermal expansion coefficient $\alpha = 25.6 \pm 0.9$ ppm/K was determined.

### B. Specific heat measurements

Specific heat measurements were performed on a single crystal of $Sc_3Mn_3Al_7Si_5$ with a mass of 8.95 mg using a Quantum Design Physical Property Measurement System (PPMS) for temperatures ranging from 1.8 K to 220 K.

### C. Inelastic Neutron Scattering measurements

INS measurements were performed using the time of flight instrument SEQUOIA at the Spallation Neutron Source at Oak Ridge National Laboratory [17,18]. We used the coarse resolution Fermi chopper running at a frequency of 180 Hz to select neutrons with an incident energy $E_i$=60 meV. Approximately 4 g of $Sc_3Mn_3Al_7Si_5$ single crystals were co-aligned for measurements in the HHL scattering plane and attached to an aluminum sample mount using aluminum wire. The assembly was subsequently placed in a bottom loading closed cycle cryostat with a base temperature of 5 K. To reduce background, a cadmium sheet was used to shield the sample mount and the inside of the cryogenic system from the incident beam and from neutrons scattered from the sample. Data were obtained at temperatures of 5 K, 100 K, and 250 K, and a good sampling of $Q$-space was obtained by rotating the sample between -10° and 190° in 1° steps in the HHL-plane, where the zero angle starting from (0, 0, 1) is perpendicular to the incident beam. An empty sample holder was measured separately at these same temperatures, providing a background to permit the isolation of the sample signal.

The powder averaged dynamic structure factor, $S(Q, E)$, with neutron wave vector transfer $Q$ and neutron energy transfer $E$, was obtained from the measured count rates using the standard software MantidPlot [19]. The contribution of the empty sample holder were subtracted using Mslice/DAVE [20]. The neutron-weighted generalized phonon density of states (GDOS) was obtained from $S(Q, E)$ for wave vectors $Q$ between 1.5 Å$^{-1}$– 4.5 Å$^{-1}$, using the following equation [21–24],

$$GDOS = \left[1 - e^{\frac{-E}{k_B T}}\right] \int \frac{E}{Q^2} S(Q, E) dQ \quad\ldots\ldots\ldots\ldots\ldots\ldots\ldots\ldots(1)$$

where $\left[1 - e^{\frac{-E}{k_B T}}\right] = (n(E) + 1)^{-1}$ accounts for the phonon Bose-Einstein statistics, $k_B$ is the Boltzmann constant, and $T$ is the temperature. No correction was made for the Debye-Waller factor.

### D. Computational methods

The Kohn–Sham density-functional theory (DFT) calculations were performed with the projector augmented wave method [25] as implemented in the Vienna *ab initio* simulation package [26]. The Perdew-Burke-Ernzerhof form of the generalized gradient approximation [27,28] was used to describe the exchange-correlation energy. All self-consistent calculations were performed with a plane wave kinetic energy cutoff of 500 eV on a 7×7×6 Γ-centered Monkhorst-Pack k-mesh. All geometrical structures were fully relaxed until the residual forces on each atom were less than 0.01 eV/Å and the total energy variation was less than $1.0 \times 10^{-6}$ eV. The real-space interatomic force constants were calculated with a 2×2×2 supercell containing 288 atoms within the harmonic approximation via the density functional perturbation theory method [29] as implemented in the PHONOPY code [30]. The calculated mechanical properties are shown in Table 1. To compare with the INS data, a simulation of the generalized phonon density of states (GPDOS) has been carried out by summing the partial phonon density of states (PhDOS$_i$) associated with a specific atom *i* weighted by the neutron weighting factor $\frac{\sigma_i}{M_i}$ which is the ratio of the atomic neutron scattering cross section $\sigma_i$ and mass $M_i$ [23,31]:

$$GPDOS = \sum_i \frac{\sigma_i}{M_i} PhDOS_i \quad \ldots\ldots\ldots\ldots\ldots\ldots\ldots\ldots\ldots\ldots(2)$$

Here, we use the abbreviation GPDOS for the generalized phonon density of states from this simulation to distinguish it from the measured generalized phonon density of states GDOS.

We convoluted GPDOS with a Gaussian approximation to the energy resolution of SEQUOIA with a full peak width at half maximum (FWHM) given by [17]:

FWHM=$(1.435*10^{-6}$ meV$^{-2})\cdot E^3+(0.0003885$ meV$^{-1})\cdot E^2-0.07882\cdot E+4.295$ meV$\ldots\ldots\ldots$(3)

The calculations of the multiphonon and multiple scattering in $Sc_3Mn_3Al_7Si_5$ were performed for a fixed temperature of 250 K data using the Multiphonon density of states tools [32]. In order to establish a common set of units, the one-phonon DOS was also computed.

## III.　　RESULTS AND DISCUSSION

Measurements of co-aligned single crystals allow us to compare directly the calculated and measured phonon dispersions. The energy dependence of the scattering at 5 K along the [H, H, 0] direction is shown in Fig. 2, where the data are summed over -1 ≤ L ≤ 1. Comparing the scattering from the sample holder plus sample (Fig. 2(a)) to the sample holder alone (Fig. 2(b)), it is clear that the sample holder makes a large contribution to the scattering. The difference of the two signals is presented in Fig. 2(c). There is reasonable correspondence to the computed phonon dispersions (Fig. 2(d)), with both the calculated and measured phonons show the same dispersion for the acoustic band, and we see the same broad band of phonons found between ~20 meV -50 meV. However, the scattering at the lowest energies where the magnetic scattering is most likely to be found is very weak, with no evidence for the acoustic phonons expected to emanate from the [-3, -3, 0] and [-1, -1, 0] Bragg points, let alone any magnetic scattering that is likely even weaker. In order to compare the INS data directly to the calculated phonon DOS, and to compensate for the small scattering intensity observed at low momentum transfer, we will predominantly consider the data averaged over all momentum transfer directions.

The energy and momentum dependencies of the total measured scattering of the sample are determined from the difference between measurements carried out on the sample plus sample holder together (Fig. 3(a)), and a separate measurement of the sample holder itself (Fig. 3(b)). The energy and wave vector dependencies of the difference signal are presented in Fig. 3(c). While the sample holder and sample itself contribute roughly equally to the overall scattering, the main features of the sample scattering in Fig. 3(c) are qualitatively what is expected. In particular, the intensity of the broad feature near 20 meV increases with increasing wave vector, as expected for phonons. For comparison of the magnitudes of the different contributions, the data from Figs. 3(a-c) are integrated over the wave vector magnitudes 1.5 Å$^{-1}$ ≤ $Q$ ≤ 4.5 Å$^{-1}$, and plotted as functions of energy (Fig. 3(d)). As well, the comparison of the sample scattering to the results of the *ab initio* calculation of the phonon DOS requires normalization by a scale factor. First, the elastic component of the scattering, which is broadened by the instrumental resolution, is replaced by the phonon DOS of a Debye model [33]. Next, the experimental data are scaled so the integrated intensities for energies 0 meV ≤ $E$ ≤ 50 meV are equal to that of the simulation. This normalization procedure shows that there is excellent agreement between the experimental and simulated phonon DOS, both in terms of the peak energies and the intensities. The computed multiple scattering and multiphonon scattering DOS are also presented in Fig. 3(d) after normalization to the one-phonon DOS using the Multiphonon density of states tool [32]. Given their very weak intensities, it is clear that multiple scattering and multiphonon scattering make only minimal contributions to the overall scattering. Only a small difference between the measurements and the simulations is visible near 23 meV. Finally, we note that the sample scattering lies below the simulated phonon DOS in the energy range 10 meV ≤ $E$ ≤ 20 meV. This is due to a signal from a brass fastener that is present in measurements of the empty sample holder, but absent in the sample plus sample holder measurement. This can be rationalized by noting that the phonon DOS of Cu, the main constituent of brass, has a peak near ~18 meV [34] that could explain the over subtraction in this range of energies. The energy-integrated intensities of the different quantities presented in Fig. 3(d) provide a good measure of their relative contributions to the overall scattering. The percentage of the total scattering derived from the 250 K data are as follows: the sample holder contribution ~59%, the total sample scattering ~41%, consisting of one-phonon scattering (36%), multiphonon scattering (4%), and multiple scattering (1%). We note that relative to the total scattering the multiphonon scattering is even smaller at lower temperatures [35]. The main conclusion that may be drawn from Fig. 3 is that the primary sample scattering above 10 meV is one-phonon scattering, and that there is no appreciable scattering from sources other than phonons in this range of energies.

Given that phonons are bosons, it is expected that the temperature dependence of the $S(Q, E)$ follows the Bose occupation factor as in Eq. 1. Deviations would suggest the presence of other factors, such as a phase transition, magnetic contributions, or phonon anharmonicity. The GDOS, where $S(Q, E)$ is modified by the Bose occupation factor (Eq. 1) is presented at 5 K, 100 K, and 250 K in Fig. 4. While the GDOS curves are offset from each other for clarity, there is little difference among them, in terms of the peak energies and breadths, that would indicate appreciable phonon softening or anharmonicity. No new features have been found in the GDOS over the measured temperature range that might be ascribed to scattering beyond that of the one-phonon DOS, at least for energies above ~10 meV.

The Weiss temperature $\theta_W$=-38 K [10] provides an approximate measure of the energy scale for magnetic excitations in $Sc_3Mn_3Al_7Si_5$, and so we sought evidence for them in the scattering below 10 meV. In order to directly compare the scattering at different temperatures, the raw data are corrected by the Bose factor and the resulting energy dependencies of the imaginary part of the dynamical susceptibility $\chi''(E)$ are presented in Fig. 5, where the data have been summed over different ranges of wave vectors. The data with 0.5 Å$^{-1}$ ≤ $Q$ ≤ 1 Å$^{-1}$ are more intense at 5 K (Fig. 5(a)) than at 250 K (Fig. 5(b)), although for those with large wave vectors, for instance 2 Å$^{-1}$ ≤ $Q$ ≤ 2.5 Å$^{-1}$, there is much less difference between the two temperatures. The difference between the two data sets is presented in Fig. 5(c), indicating that at the lowest $Q$ range there is additional scattering present at 5 K that is absent at 250 K. This additional scattering meets basic requirements that identify it as magnetic scattering. Namely, it is strongest at lowest $Q$, due to the magnetic form factor. It is strongest at low temperatures, where the system is closest to a $T$=0 magnetic transition. Finally, the scattering (Fig. 5(c)) is broad and quasielastic, as is often found in quantum critical systems [36–38]. Figure 5(a) indicates that the proposed magnetic scattering coexists with phonon scattering on this range of energies. Isolating the proposed magnetic signal will be challenging given that it is very weak, almost at the limit of sensitivity for this measurement.

More information about the phonons and their dispersions is available from the *ab initio* calculations. Figure 6(b) shows the phonon dispersion along the Γ-M-K-Γ-A-L-H-A path through the Brillouin zone (Fig. 6(a)). All of the phonon branches have energies exceeding 10 meV at the Brillouin zone boundaries, and a significant portion of the phonon DOS is found between ~10 – 50 meV in qualitative agreement with the measured DOS in Figs. 3 and 4. Since the atomic neutron scattering cross section weights the phonon modes by $\sigma_i/M_i$ (Eq. 2), it is possible to define their partial DOS, which are their respective contributions to the overall GPDOS. In this way, the corresponding total and partial phonon DOS for each atom can be determined (Fig. 6(c)). The partial phonon DOS shown in Fig. 6(c) shows that Sc dominates the total neutron-weighted phonon DOS. This is due to the large neutron scattering cross section of Sc (23.5 barn) [39]. Conversely, the contribution from Mn is rather small, due to its small neutron scattering cross section (2.15 barn) [39].

Having established the phonon DOS for $Sc_3Mn_3Al_7Si_5$ through INS and DFT, we use it to determine the phonon contribution to the specific heat [33]. The energy dependence of the neutron weighting factor $\sum_i \frac{\sigma_i}{M_i}$ is determined from *ab initio* calculations through the ratio of neutron weighted GPDOS and the pure phonon DOS (PDOS) (Fig. 7(a-b)). The normalized experimental PDOS is determined by dividing the GDOS measured at 5 K (Fig. 7(c)) by the weighting ratio (Fig. 7(b)). The result is shown in Fig. 7(d), where the resolution broadened elastic peak has been replaced by the DOS of a Debye model to account for the acoustic phonons. This is used to compute the corresponding harmonic phonon contribution to the specific heat for all temperatures [33].

The measured specific heat $C_P$ has several components

$$C_P = C_{phonon} + C_{dilation} + C_{electron} \quad \text{...........................(4)}$$

where $C_{phonon}$ is the contribution from the phonons, $C_{dilation}$ is the specific heat due to the increase of the lattice constant with increasing temperature, and $C_{electron}$ represents any remaining part of the specific heat, which we associate with the electronic and magnetic contributions to the specific heat. $C_{phonon}$ and $C_{dilation}$ are defined by the following equations [33,40]

$$C_{phonon} = k_B \int_0^\infty D(E) dE \left(\frac{E}{k_B T}\right)^2 \frac{e^{\left(\frac{E}{k_B T}\right)}}{\left[e^{\left(\frac{E}{k_B T}\right)} - 1\right]^2} \quad \text{.....................(5)}$$

$$C_{dilation} = BV\alpha^2 T \quad \text{..........................................(6)}$$

where $D(E)$ is the PDOS determined from the INS measurements (Fig. 7(d)), $B$=124 GPa is the isothermal bulk modulus determined from *ab initio* calculations (Table 1), $V$ is the specific volume, and $\alpha = 25.6 \pm 0.9$ ppm/K is the volumetric thermal expansion coefficient. The energy cutoff in the integral of Eq. 5 is set at 55 meV, beyond which the phonon DOS (Fig. 6(b) and Fig. 7(d)) becomes very small. The dilational component of the specific heat is determined to be $C_{dilation}/T = (13.3 \pm 0.9)$ mJ·mol$^{-1}$·K$^{-2}$.

The measured specific heat $C_P$ of Sc$_3$Mn$_3$Al$_7$Si$_5$ is plotted in Fig. 8 along with $C_{phonon}$ and $C_{dilation}$. The remainder of the specific heat defines the combined magnetic and electronic specific heat $C_{electron} = C_P - (C_{phonon} + C_{dilation})$. The dilation component is very small over the entire range of temperatures up to 230 K. $C_{electron}$ is found to be approximately linear at temperatures above ~125 K with a fitted slope of $\gamma = (56 \pm 1)$

mJ·mol$^{-1}$·K$^{-2}$. Our previous analysis of the specific heat [10] showed a pronounced increase in $C_P/T$ below ~10 K, indicating that the magnetic scattering is small or absent at higher temperatures, where the specific heat is well described by a Debye model plus a somewhat larger linear term of 80 mJ·mol$^{-1}$·K$^{-2}$, representing the electronic specific heat. Figure 8 shows that the sum of $C_{\text{phonon}} + C_{\text{dilation}} + \gamma T$ provides an excellent description of the measured specific heat $C_P$, not only in terms of its magnitude, but also with respect to the details of its temperature dependence over the entire range of temperatures from 30 K to 220 K.

Table 2 compares the Sommerfeld constant of Sc$_3$Mn$_3$Al$_7$Si$_5$ to those of other Mn-based metals. It is clear that the value $\gamma = (56 \pm 1)$ mJ·mol$^{-1}$·K$^{-2}$ is among the largest of the values found in compounds like CaMn$_2$Al$_{10}$ [41], Ti$_4$MnBi$_2$ [42], and Mn$_2$Sb [43], while being significantly larger than more weakly correlated metals like MnAlGe [44], MnP [45], or $\alpha$-Mn itself [46]. This comparison highlights the importance of electronic correlations, presumably originating with the Mn $d$-electrons. In an insulator, strong Coulomb interactions lead to localized magnetic moments, where the electrons occupy states that are determined by the crystal electric field and spin orbit coupling. The Curie-Weiss effective moment for Sc$_3$Mn$_3$Al$_7$Si$_5$ is 0.5 $\mu_B$/Mn [10], indicating that the Mn moments in Sc$_3$Mn$_3$Al$_7$Si$_5$ are rather itinerant. In this case, the initially localized $d$-orbitals are hybridized into broad bands. The residual $d$-electron character from the Mn atoms results in a moderate localization of the itinerant states that leads to the mass enhancement evident from the large Sommerfeld coefficient of Sc$_3$Mn$_3$Al$_7$Si$_5$. Given this intermediate degree of itineracy and the corresponding correlations, the broad and weak magnetic scattering in Sc$_3$Mn$_3$Al$_7$Si$_5$ suggests a comparison to INS measurements carried out in the paramagnetic phases of transition metal ferromagnets [47–49]. In those cases, the scattering originates from critical diffusive fluctuations and overdamped spin waves, leading to scattering that is very broad with respect to energy and wave vector, as is nominally consistent with the proposed magnetic scattering in $Sc_3Mn_3Al_7Si_5$.

## IV. CONCLUSIONS

We have presented here the results of INS measurements in a metallic kagome system $Sc_3Mn_3Al_7Si_5$. Phonon scattering is dominant over much of the energy range, and good correspondence is demonstrated between the measured and *ab initio* calculations of the phonon DOS. As well, the specific heat determined using the measured DOS is in quantitative agreement with the measured temperature dependence of the specific heat. Given the itinerant character of the Mn magnetism in $Sc_3Mn_3Al_7Si_5$ that is revealed by the reduced Curie-Weiss moment, it is expected that any magnetic scattering will be weak. Indeed, we isolated a magnetic signal that was present only at low temperature, distinguishable from the stronger phonon scattering by its wave vector dependence. This signal is broad and quasielastic, as expected for metallic paramagnets with itinerant moments. In particular, these experiments provide no evidence at these temperatures and energies for the fractionalized excitations that have been found in some insulating kagome systems. Further study will be needed to understand the exact nature of the weak magnetic excitations that are reported here.

**ACKNOWLEDGEMENTS**

We thank B. Patrick for assistance in carrying out the single crystal XRD measurements. X. Y. Li and M. C. Aronson acknowledge support from NSF under Award No. NSF-DMR-1807451. A. M. Hallas acknowledges support from NSERC of Canada and the CIFAR Azrieli Global Scholars Program. This research was undertaken thanks in part to funding from the Max Planck-UBC-UTokyo Centre for Quantum Materials and the Canada First Research Excellence Fund, Quantum Materials and Future Technologies Program. B.-T.Wang and P.-F. Liu acknowledge financial support from the Natural Science Foundation of China (Grant No. 12074381). C. Broholm was supported as part of the Institute for Quantum Matter, an Energy Frontier Research Center funded by the U.S. Department of Energy, Office of Science, Basic Energy Sciences under Award No. DE-SC0019331. A portion of this research used resources at the Spallation Neutron Source, a DOE office of Science User Facility operated by the Oak Ridge National Laboratory.

Figures:

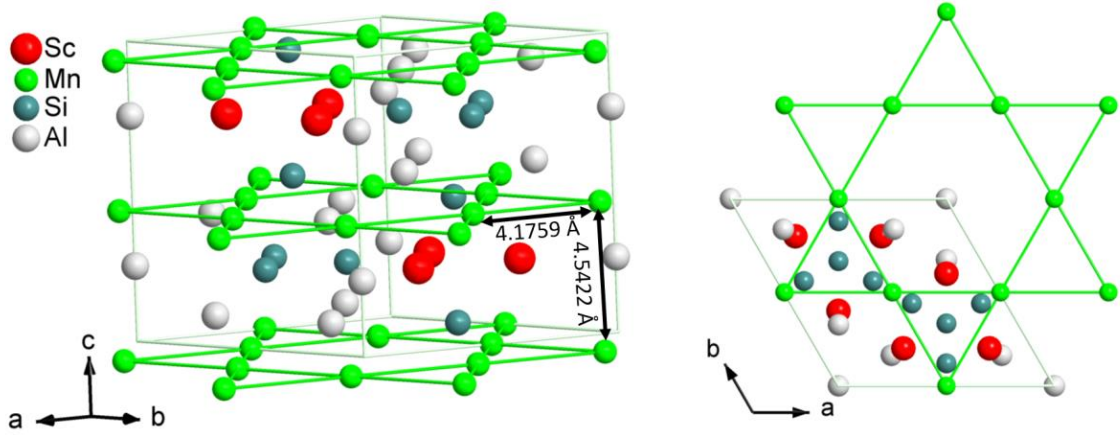

FIG. 1: The hexagonal crystal structure of $Sc_3Mn_3Al_7Si_5$ with space group $P6_3mmc$ (No.194), where the Mn atoms (green spheres) form kagome lattices in the $ab$ plane.

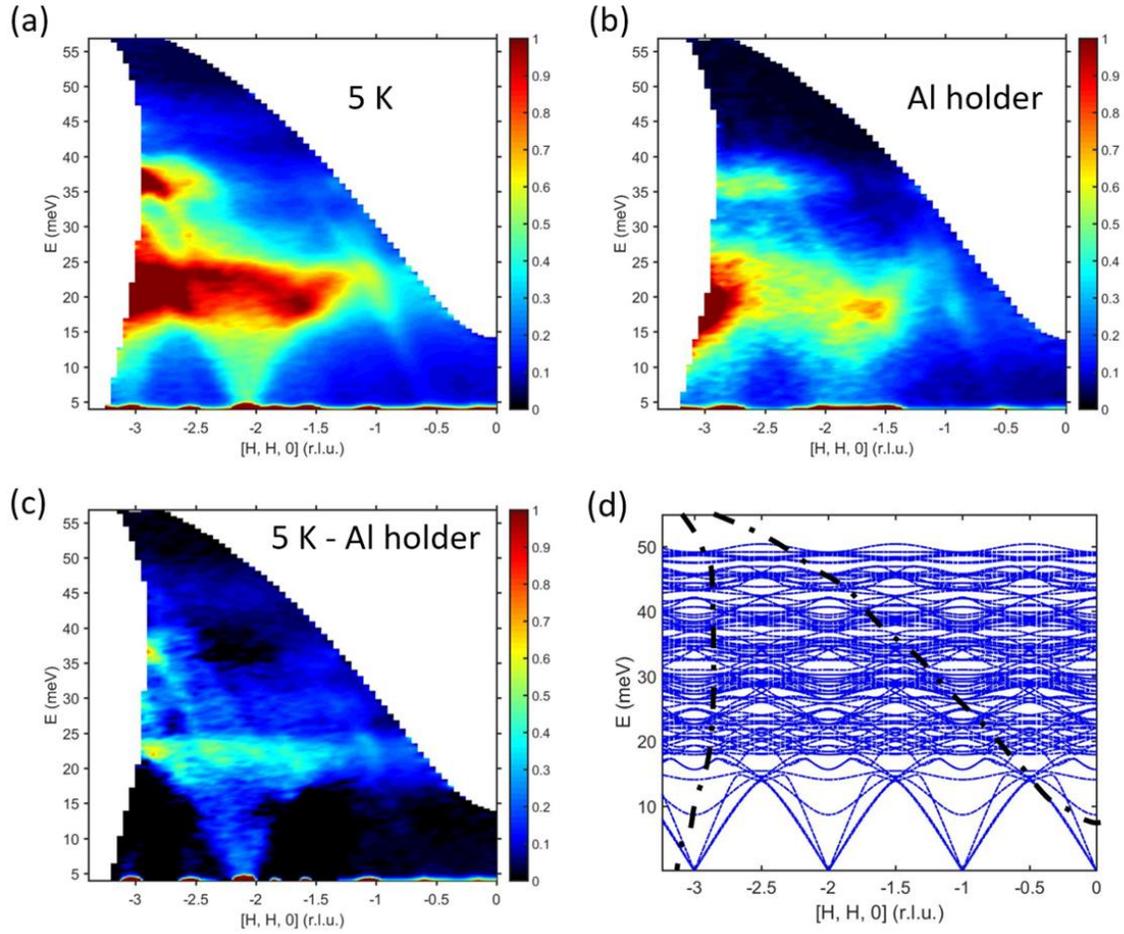

FIG. 2: Energy dependence of the 5 K scattering in the [H, H, 0] plane, where the data are summed -1 ≤ L ≤ 1. (a) Sample plus sample holder. (b) Sample holder only. (c) Difference of (a) and (b), nominally representing the sample scattering. (d) The computed phonon dispersions.

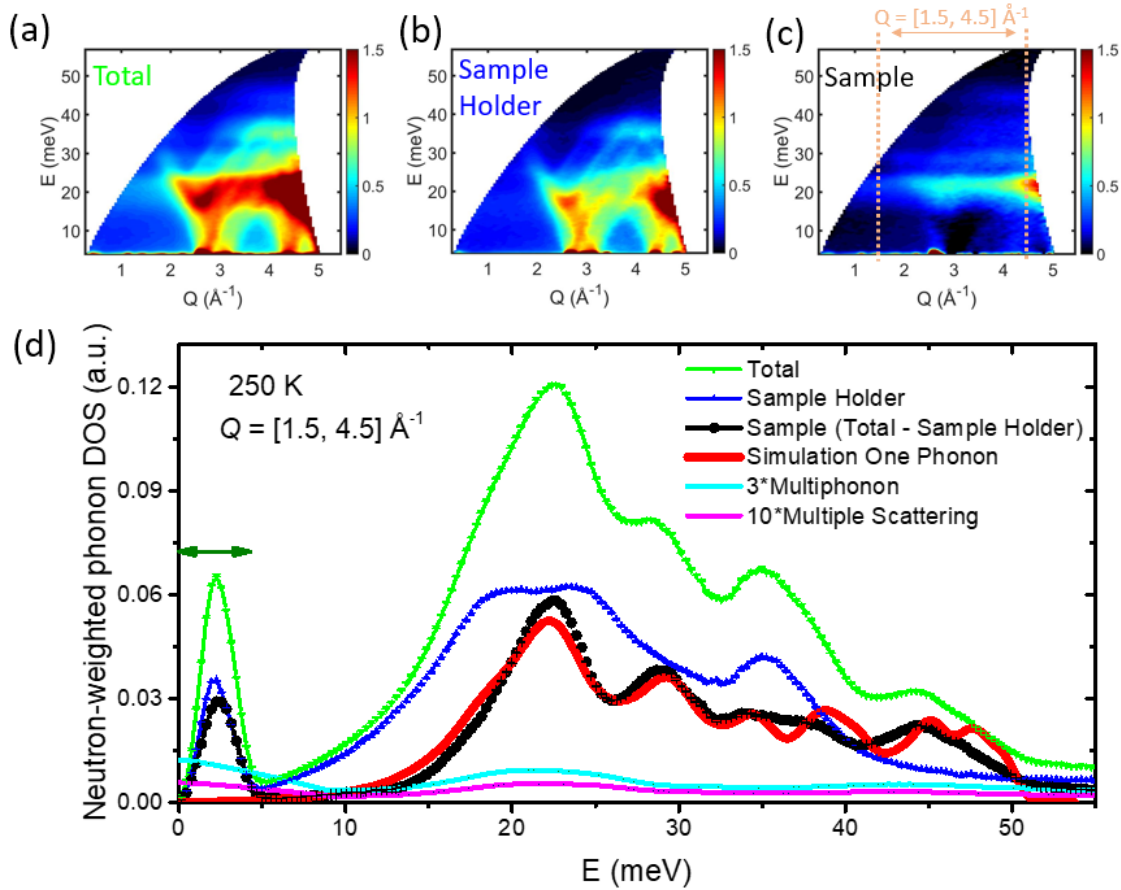

FIG. 3: Energy and momentum dependence of the powder averaged data. (a) Sample plus sample holder (Total). (b) Sample holder only. (c) Difference of (a) and (b), nominally representing the sample scattering. (d) Neutron-weighted phonon DOS of $Sc_3Mn_3Al_7Si_5$. The INS data were measured at SEQUOIA with incident neutron energy $E_i = 60$ meV. The total sample plus sample holder (green) and the empty sample holder (blue) were measured separately at 250 K, and the signal from the sample (black) is given by their difference. The sample signal is compared to the calculated one-phonon DOS (red), with the former normalized to the energy integrated intensity of the latter yielding excellent agreement. The scattering from the sample holder is well described by the phonon DOS of aluminum, apart from a contribution from a brass fastener material that leads to over subtraction below 20 meV, where the phonon DOS of Cu has a peak at ~ 18 meV. The computed multiphonon

(cyan, multiplied by a factor of three) and multiple scattering (magenta, multiplied by a factor of ten) contributions are also presented. Both are very weak, since their energy integrated intensities are respectively 4% and 1% of the total measured intensity, which is dominated by the Al background of the sample holder (59%) and the one-phonon density of states of the sample (36%). The peaks at ~2 meV results from the interplay of two effects: the inclusion of the $E/Q^2$ factor in the GDOS (Eq. 1), convoluted with the energy resolution of the spectrometer (green arrows). The statistical error bars are smaller than the size of the data points.

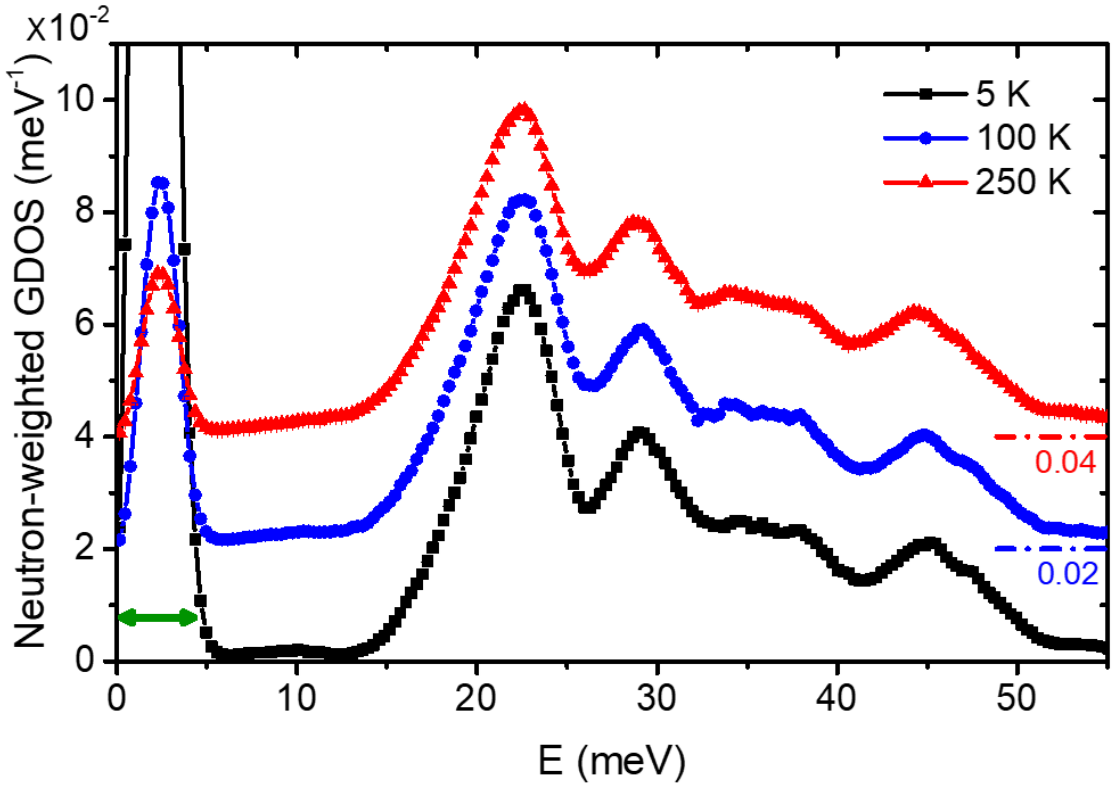

FIG. 4: Neutron-weighted generalized phonon density of states GDOS of $Sc_3Mn_3Al_7Si_5$ from measurements at the indicated temperatures on SEQUOIA. The elastic scattering broadened by the instrumental resolution leads to the broad peaks below ~5 meV, where the green double arrows show the instrumental resolution. The 100 K and 250 K data are offset by 0.02 meV$^{-1}$ and 0.04 meV$^{-1}$ along the y-axis, respectively, as indicated by the dashed-dotted lines. The statistical error bars are smaller than the size of the data points. The peaks at ~2 meV results from the interplay of two effects: the inclusion of the $E/Q^2$ factor in the GDOS (Eq. 1), convoluted with the energy resolution of the spectrometer (green arrows).

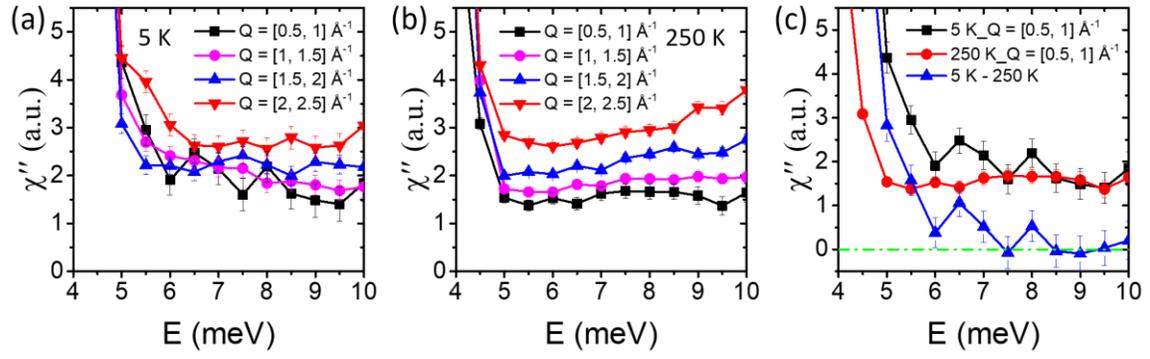

FIG. 5. Imaginary part of the dynamical susceptibility χ"(E) reported for different ranges of wave vector $Q$ at 5 K (a) and 250 K (b). (c) Comparison of χ"(E) for 0.6 Å$^{-1}$ ≤ $Q$ ≤ 1 Å$^{-1}$ at 5 K (black), 250 K (red) and their difference (blue).

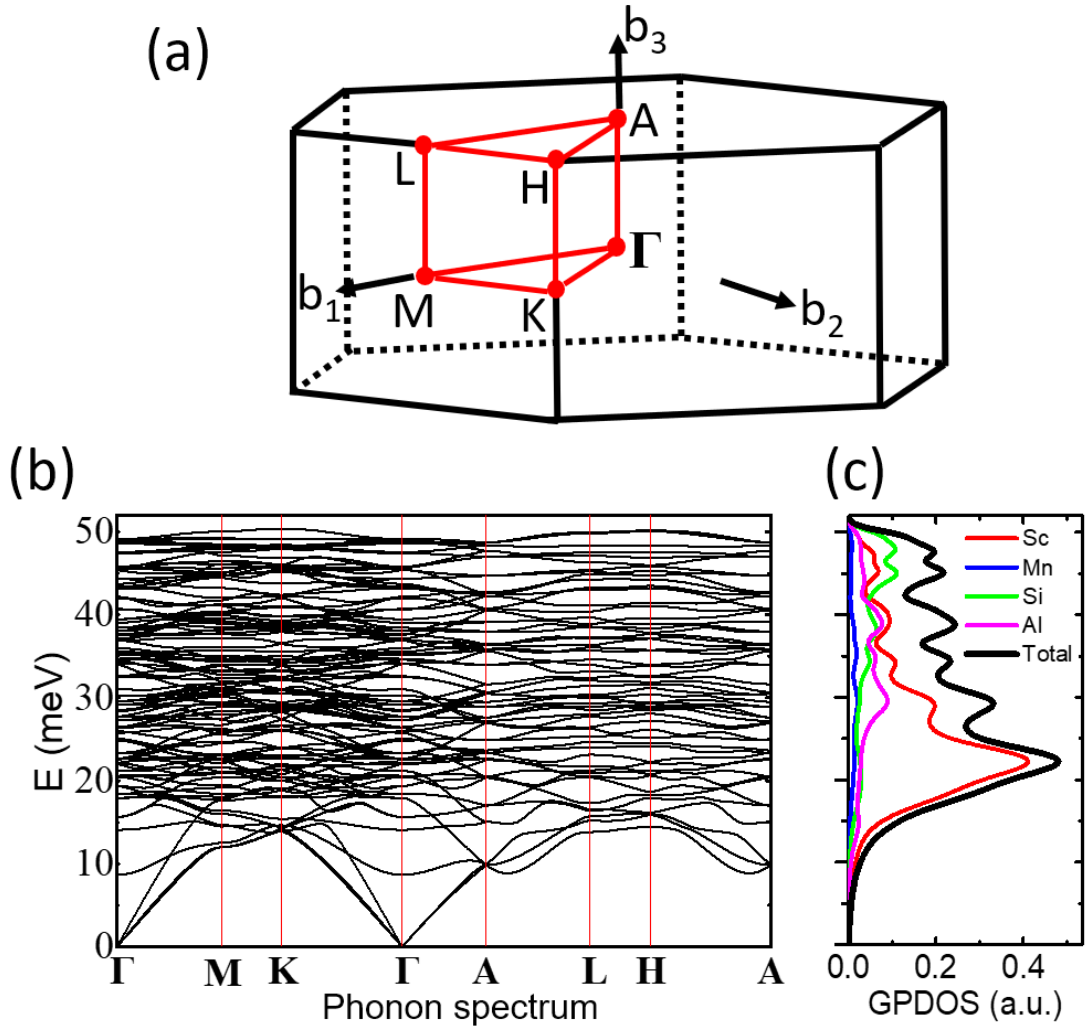

FIG. 6: *ab initio* calculations of the phonon spectrum of $Sc_3Mn_3Al_7Si_5$, along Γ-M-K-Γ-A-L-H-A. (a) The Brillouin zone. (b) The calculated phonon spectrum and (c) generalized neutron weighted phonon density of states GPDOS. The contributions to the phonon GPDOS from different atoms are shown separately, as well as the total. The effects of the neutron scattering cross sections for each atom are included in the GPDOS calculation (Eq. 2), which can be directly compared with the GDOS extracted from the INS measurement (Fig.3).

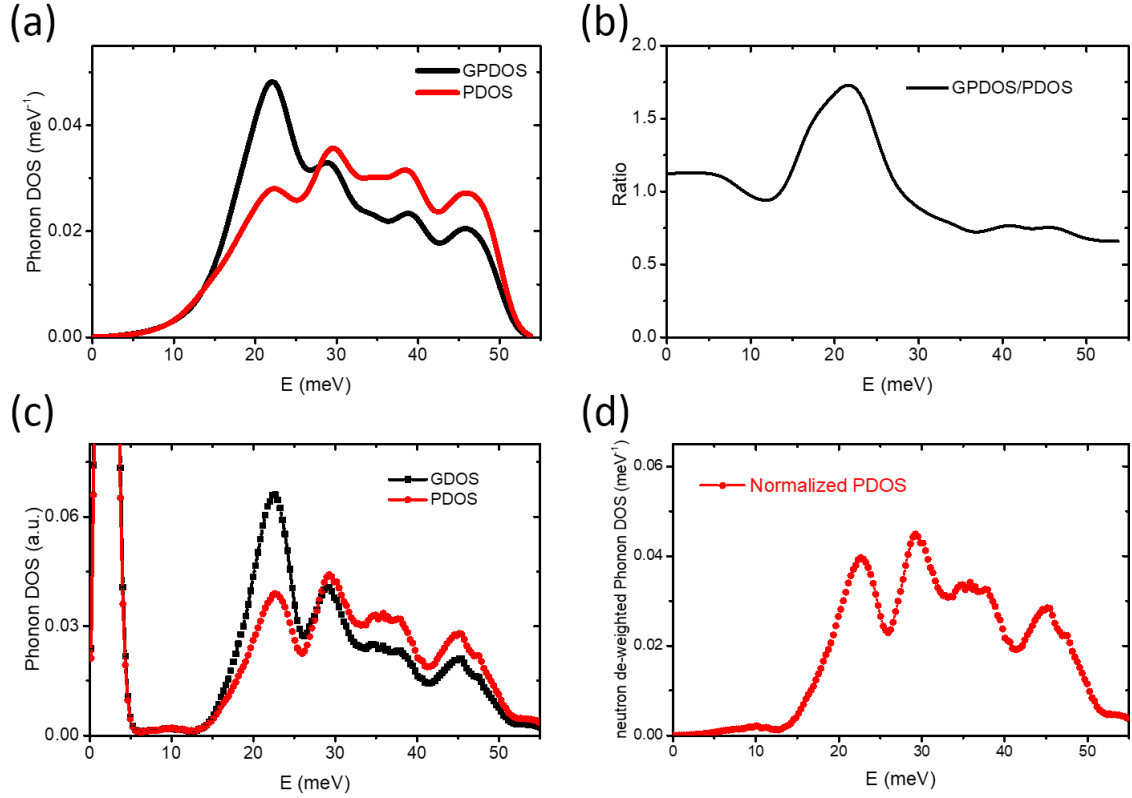

FIG. 7: Phonon DOS of $Sc_3Mn_3Al_7Si_5$. (a) The *ab initio* calculations of the neutron-weighted GPDOS and the pure phonon DOS (PDOS). (b) The ratio between the GPDOS and PDOS, which is used to convert the INS data at 5 K into the neutron de-weighted experimental PDOS (c). The normalized experimental PDOS, where the statistical error bars are smaller than the data points. (d) is obtained by subtracting out the resolution broadened elastic and thermal diffuse scattering peak and adding back in the DOS of a Debye model.

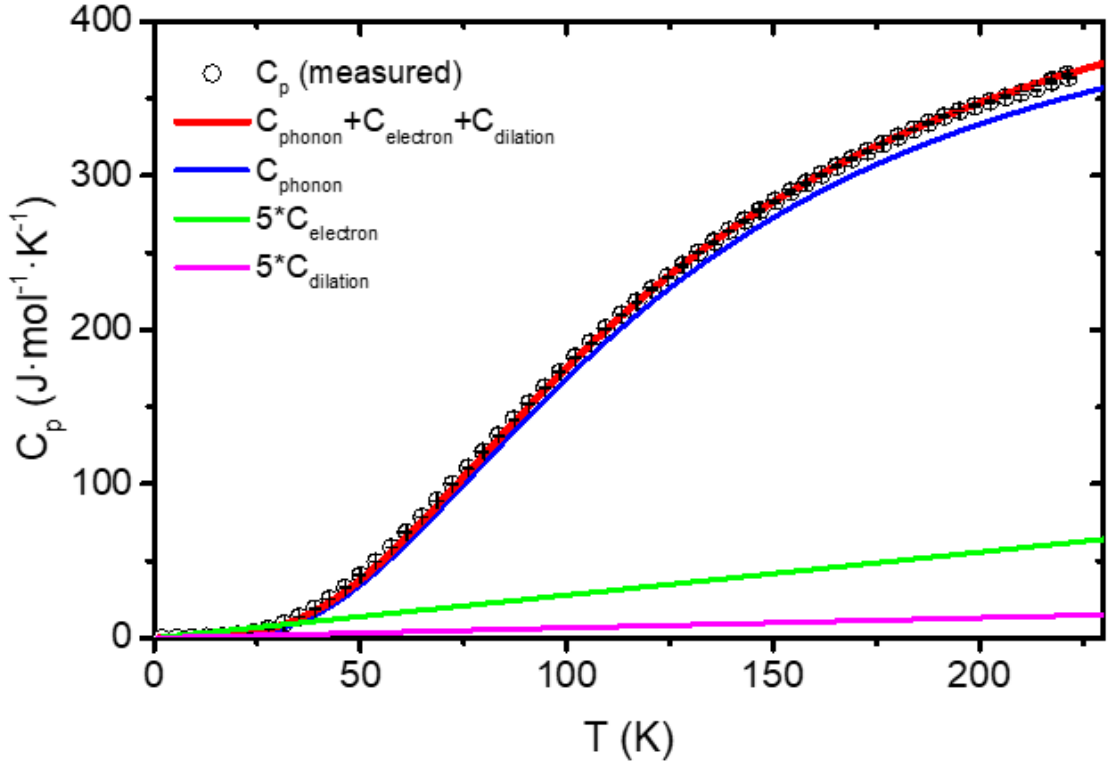

FIG. 8: The measured specific heat $C_P$ of $Sc_3Mn_3Al_7Si_5$ (open black circles). The blue line indicates the integral over the phonon DOS, which is extracted from INS measurements, as described in the text. The lattice dilation component $C_{dilation} = [13.3\pm0.09\ mJ\cdot mol^{-1}\cdot K^{-2}]T$ (magenta line, multiplied by a factor of five) is determined from the measured temperature dependence of the lattice constant, and the electronic component $C_{electron} = [56\pm1\ mJ\cdot mol^{-1}\cdot K^{-2}]T$ (green line, multiplied by a factor of five) is inferred by subtracting $C_{phonon}$ and $C_{dilation}$ from the measured specific heat $C_P$. The red line indicates the sum of the specific heat $C_{phonon}$ that is calculated from the phonon density of states, the lattice dilational component $C_{dilation}$, and the electronic component $C_{electron}$.

Table 1: Average mechanical properties for polycrystalline Sc$_3$Mn$_3$Al$_7$Si$_5$ calculated by *ab initio* simulations.

| | |
|---|---|
| Bulk modulus K (GPa) | 124.205 |
| Shear modulus G (GPa) | 100.361 |
| Young's modulus E (GPa) | 237.196 |
| P-wave modulus (GPa) | 258.020 |
| Poisson's ratio v | 0.182 |
| Bulk/Shear ratio | 1.238 |
| Pugh Ratio | 0.809 |
| Cauchy Pressure (GPa) | -36.482 |
| Universal Elastic Anisotropy | 0.013 |
| Chung-Buessem Anisotropy | 0.001 |
| Isotropic Poisson's Ratio | 0.181 |
| Longitudinal wave velocity (in m/s) | 8143.531 |
| Transverse wave velocity (in m/s) | 5080.700 |
| Average wave velocity (in m/s) | 5597.972 |
| Debye temperature (in K) | 676.748 |

Table 2: The Sommerfeld constant $\gamma$ for several Mn-based metals. The molar unit is the formula unit throughout.

| | mJ·mol$^{-1}$·K$^{-2}$ |
|---|---|
| Sc$_3$Mn$_3$Al$_7$Si$_5$ (This work) | 56 |
| HfMnGa$_2$ [50] | 33 |
| CaMn$_2$Al$_{10}$ [41] | 40 |
| Ti$_4$MnBi$_2$ [42] | 57 |
| MnP [45] | 9.65 |
| Mn$_2$Sb [43] | 71.4 |
| MnSi [51] | 36.7 |
| SrMnBi$_2$ [52] | 36.5 |
| MnAlGe [44] | 8.9 |
| MnZnSb [44] | 11.3 |
| $\alpha$-Mn [46] | 0.118 |